# More General Soliton Solution for Vectorial Bose-Einstein Condensate

P. Sakthivinayagam[1*]

Department of Physics, PSG College of Arts & Science (Autonomous), Coimbatore, Tamil Nadu, India– 641 014.

Manuscript No:

## Abstract

WE derive exact and more general solutions of the two coupled Gross-Pitaevskii equation with suitable parameters by demonstrating two analytical methods. In the first method, equations are analyzed and inferred some of their mathematical and physical properties, which are then used to derive the exact stationary solutions. In the second method, we demonstrate the Darboux transformation method and construct exact and more general soliton solutions for the Gross-Pitaevskii equation (NLS equation with external potential term). We have proved that the solutions were more general one by showcasing all kinds of soliton pairs by maneuvering the parameters suitably.

1. P. Sakthivinayagam, Assistant Professor (SG), PSG College of Arts & Science, Department of Physics, Coimbatore, Tamil Nadu, India (Corresponding author, mobile No: 9655755113, e-mail: (sakthivinayagam@psgcas.ac.in).

## I. INTRODUCTION

EVER since our birth, we humans interact with a lot of physical objects. From casual observations, we note that object, or strictly speaking in the language of science, matter is most commonly found in 3 different forms of state. When we venture deep and try to understand the crux of this behavior, we see that state is a function of the energy of the matter. By altering the temperature of the system, we can traverse between the states. Ambitious scientists tried to push the systems to high temperatures, just to understand the dynamics, but in that process discovered the plasma state. It's a state where the temperatures are steeply inflated, and the atoms are super excited. This discovery attracted a huge share of scientists into plasma physics and gravitated their attention, solely because of its intense application due to its enormous abundance in the universe. Maneuvering our brainstorming, we ask ourselves a question of, how will the system behave at ultra-low temperature? Introduced to the world by Bose and Einstein during the 1920s, achieving BEC was problematic due to technological constrain. It took seven decades to achieve BEC and was accomplished by Cornell and Weiman. By cooling atoms to near zero Kelvin, Bose-Einstein condensate was bought to life [1, 2]. It is a complex quantum phenomenon where the particle nature is reformed into a matter-wave when the atoms are super cooled. BECs are achieved by cooling down boson gases like the ensemble of Helium 4 atoms. Now, what are Bosons? To answer that question, we go back to quantum mechanics and try to understand how it predicts the presence of two very special particles. They are the Bosons and Fermions. To understand what makes some particles bosons and fermions, we need to look into their wavefunctions.

Wavefunction $\psi(r)$ of any system gives us valuable information about the system. When we square the wavefunction, we end up with $\psi(r)^2$ which dictates the probability distribution function of the particles in the system. Now, probability distribution gives us a range over which we can detect the particles. We can create wavefunction for any number of particles in the system and the corresponding probability distribution function can be obtained. Now assume a system with two identical and indistinguishable particles. By this, we understand that both systems have the same wave function. As the particles are indistinguishable, we will not be able to differentiate them if they are swapped. This means the probability distribution for two particles in any arbitrary points say $r_1$ and $r_2$ are the same [3,4].

Mathematically, $\psi(r_1, r_2)^2 = \psi(r_2, r_1)^2$, is the probability distribution function of the particles. By taking the square root on both the sides, we get $\psi(r_1, r_2) = \pm \psi(r_1, r_2)$. Therefore, after the swap, the wave function remains the same but introduces a phase difference. One class of solutions we have here is $\psi(r_1, r_2) = +\psi(r_1, r_2)$ and the other class of solutions is $\psi(r_1, r_2) = -\psi(r_1, r_2)$. In experimental observations, we have detected different particles obeying either of the above equations. This validates our assumption of the presence of indistinguishable particles. The particles that obey equation $\psi(r_1, r_2) = +\psi(r_1, r_2)$ are homologous to each other and are called the Bosons, and the particles that obey $\psi(r_1, r_2) = -\psi(r_1, r_2)$ are called the Fermions. The fundamental understanding from the above statement is that particles in bosons have symmetric wavefunction and fermions have anti-symmetric wave function. Assume any two energy levels say $E_1$ & $E_1$, and two identical, indistinguishable particles. As in the case of bosons, both the particle can exist at the same energy level, or different levels as given by $\psi = |0>|0>$ or $\psi = |1>|1>$ or $\psi = 1/\sqrt{2} \ |0>1> + |1>0>$. But Fermions can only take up different energy levels and their wave equation is given by $\psi = 1/\sqrt{2} \ |0>1> - |1>0>$. The negative sign is the phase difference between the two superposed states When we try to swap the latest equation the entire wavefunction will become negative like, $\psi = -1/\sqrt{2} \ |0>1> + |1>0>$. These above-mentioned qualities do not allow fermion to conglomerate in one particular state. This fundamental difference gives rise to BEC in bosons. As bosons can exist in the same quantum state, they tend to occupy the lowest energy states thus creating an opportunity to exist in BEC. According to quantum mechanics, when we decrease the temperature of bosons, they start behaving like matter waves. This phenomenon is governed by the equation $\Lambda = \frac{1}{\sqrt{T}}$ which essentially means De-Broglie wavelength increases with a decrease in temperature. When at room temperature, all the bosons are in random motion, when the temperature is decreased, the wave nature becomes prominent, and they start behaving like a wave. At ultralow temperatures, the average distance between two bosons will become less than the size of the matter waves. At one point, all the bosons amalgamate to produce a single big wave function. Further from this point, different types of cooling mechanisms like Laser cooling, Doppler cooling, Sisyphus cooling, super coil cooling, are deployed for additional condensation. It is to be noted that the effects of BECs can be observed at macroscopic levels. The theoretical model for Bose-Einstein condensate is governed by the nonlinear Schrodinger wave equation $i\psi_t + 1/2\ \psi_{xx} + \sigma|\psi|^2 = 0$. Here, $\psi = \psi_{x,t}$ is a complex function. To get Bose-Einstein condensation (BEC), it is paramount to have a confining (trapping) of the magnetic and optical potential. This is achieved mathematically by adding a potential term, $V(x, t)$ in $\psi_{x,t}$, to the NLSE. This results in the Gross-Pitaevskii equation [5-9].

In this paper we are going to investigate two component BECs which is governed by two coupled Gross-Pitaevskii (GP) equation using Darboux and Similarity transformation methods.

## II. MODEL EQUATION AND DARBOUX TRANSFORMATION

The behavior of the two component Bose-Einstein condensates that are prepared in two hyperfine states can be described at sufficiently low temperatures by the two-coupled GP equation of the following form:

$$i\psi_{1_t} + \frac{1}{2}\psi_{1_{xx}} + [V(x,t) + R_{11}|\psi_1|^2 + R_{12}|\psi_2|^2]\psi_1 = 0, \quad (1a)$$

$$i\psi_{2_t} + \frac{1}{2}\psi_{2_{xx}} + [V(x,t) + R_{21}|\psi_1|^2 + R_{22}|\psi_2|^2]\psi_2 = 0 \quad (1b)$$

Where, $\psi_{it}$ and $\psi_{ixx}$ denote respectively the first and second derivatives with respect to $t$ and $x$. The (positive) terms $R_{11,12,21,22}$ represent the attractive interactions and $V(x,t) = \omega(t)^2 x^2/2$ is the time-dependent trapping field. We first convert the system (1a, 1b) to a Manakov system via a similarity transformation has detailed in Ref [12].

$$iQ_{1t} = [-\frac{1}{2}Q_{1xx} - (|Q_1|^2 + |Q_2|^2)]Q_1 \tag{2a}$$

$$iQ_{2t} = [-\frac{1}{2}Q_{1xx} - (|Q_1|^2 + |Q_2|^2)]Q_2 \tag{2b}$$

Then, we look for analytic solutions using the Darboux transformation (DT) method. The generalized Coupled Nonlinear Schrodinger (2a, 2b) (CNLS) equations require finding a linear system of equations for an auxiliary fields $\Phi(x, t)$.

The linear system is usually written in compact form in terms of a pair of matrices as follows:

$$\Phi_x = \mathbf{U}\Phi, \tag{3a}$$
$$\Phi_t = \mathbf{V}\Phi, \tag{3b}$$

**U** and **V**, known as the Lax pair, are functionals of the solutions of the model equations. The consistency condition of the linear system $\Phi_{xt} = \Phi_{tx}$ must be equivalent to the model equation under consideration.

We find the following linear system which corresponds to the class of generalized CNLS

$$\Phi_x = \mathbf{U}_0\Phi + \mathbf{U}_1\Phi\Lambda, \tag{4a}$$

$$\Phi_t = \mathbf{V}_0\Phi + \mathbf{V}_1\Phi\Lambda + \mathbf{V}_2\Phi\Lambda^2, \tag{4b}$$

where,

The values of $\Phi$, $U_0$, $U_1$, $\Lambda$, $V_0$, $V_1$, $V_2$ are nothing but the well-known Lax Pair formalism. The exact forms of the above matrix transformation have been already given by the author Sakthivinayagam et.al., [12] where $\lambda_{1,2,3}$ is the spectral parameter. The consistency condition $\Phi_{xt} = \Phi_{tx}$ leads to $\mathbf{U}_t \cdot \mathbf{V}_x + [\mathbf{U}, \mathbf{V}] = \mathbf{0}$ which should generate the CNLS equations. The next step is to use the Darboux transformation to find the solution.

## III. DARBOUX TRANSFORMATION

A Darboux transformation is a modified gauge transformation [10]

$$\Phi[1] = T[1]\Phi = \Phi\Lambda - S_1\Phi, \tag{5}$$

where $\Phi$ and $\Phi_1$ are old and new eigenfunctions of (4), T[1] and $S_1$ is Darboux matrix and non-singular 2 x 2 matrix, the Darboux transformation (5) transforms the original Lax pair (4) into a new Lax pair,

$$\Phi[1]_X = U[1]_0\Phi[1] + U[1]_1\Phi[1]\Lambda, \tag{6a}$$

$$\Phi[1]_T = V[1]_0\Phi[1] + V[1]_1\Phi[1]\Lambda + V[1]_2\Phi[1]\Lambda^2 \tag{6b}$$

in which the matrices $U[1]_0$, $U[1]_1$, $V[1]_0$, $V[1]_1$, and $V[1]_2$ assumes the same form of $U_0$, $U_1$, $V_0$, $V_1$, and $V_2$ respectively except the field variables $q(x, t)$ and $q(x, t)^*$ have now acquired new expressions, namely $q[1](x, t)$ and $q[1](x, t)^*$ in U[1] and V[1]. Inserting equations (5) into (6) and comparing the results with (3), we find

$$U[1] = (T[1]_x + T[1]U) T[1]^{-1}, \quad (7a)$$

$$V[1] = (T[1]_t + T[1]V) T[1]^{-1} \quad (7b)$$

Plugging the expressions U[1], V[1], U, V and T[1] in Eq. (6) and equating the coefficients of various powers of $\Lambda$ both sides, we get the following relations between old and new potentials, namely

$$V_0[1] = V_0, \quad (8a)$$
$$V_1[1] = V_1 + [V_0, S_1], \quad (8b)$$
$$V_2[1] = V_2 + [V_1, S_1] + [V_0, S_1] S_1, \quad (8c)$$
$$U_0[1] = U_1 + [U_1, S_1], \quad (8d)$$
$$S_{1x} = [U_0, S_1] + [U_1, S_1] S_1, \quad (8e)$$
$$S_{1t} = [V_0, S_1] + [V_1, S_1] S_1 + [V_2, S_2] S_2 \quad (8f)$$

The eigenvalue problem (4) remains invariant under the transformation (5) provided $S_1$ satisfies all the equations (8). We assumes the general form for

$$S_1 = \left(\begin{bmatrix} S_{11} & S_{12} \\ S_{13} & S_{14} \end{bmatrix}\right) \quad (9)$$

Inserting this assumed form of $S_1$ in Equation (8d) and equating the matrix elements on both sides, we find

$$Q[1](x, t) = Q(x, t) + 2S_{12}, \quad (10)$$

$$Q[1]^*(x, t) = Q^*(x, t) - 2S_{21}, \quad (11)$$

To construct more general soliton solution of equation (1) we consider $S_1$ to be

$$S_1 = \Phi_1 \Lambda_1 \Phi_1^{-1}, \quad (12)$$

where the solution of (5) at $\Lambda = \Lambda_1$, the exact form of $\Phi_1$ and $\Lambda_1$ are given by

$$\Phi 1 = \begin{pmatrix} \psi 1(x,t) & \psi_1^*(x,t) \\ \phi_1(x,t) & \phi_1^*(x,t) \end{pmatrix}, \quad \Lambda 1 = \left(\begin{bmatrix} \lambda_1 & 0 \\ 0 & \lambda_2 \end{bmatrix}\right), \quad (13)$$

where $(\psi_1, \phi_1)^T$ is the solution for the model (1) when $\lambda=\lambda_1$ and the $\lambda=\lambda_1$ is called iso-spectral parameter. It is important to mention here that during the whole Darboux transformation the assumed matrix $S_1$ must satisfy the equation (8). Then the first iterated Darboux transformation [10] is given by $\Phi[1] = T[1]\Phi = \Phi\Lambda S[1]\Phi$ (vide Eq.((5))). If $\Phi_1$ is the solution of $\Phi$ when $\Lambda=\Lambda_1$ then it should satisfy

$$\Phi_1[1](x, t) = T[1]\Phi_1(x, t) = 0 \rightarrow S_1\Phi_1(x, t) = \Phi_1(x, t) \Lambda_1.$$
…(14)

Expressing Eq. (14) in matrix form and using Cramer's rule we can determine the exact forms of $S_{12}$ and $S_{21}$ which are given by

$$S_{12} = \frac{(\lambda_2 - \lambda_1) \psi_1(x,t)\psi_1^*(x,t)}{\psi_1(x,t)\phi_1^* - \phi_1(x,t)\psi_1^*} \quad (15a)$$

$$S_{21} = \frac{(\lambda_1 - \lambda_2) \phi_1(x,t)\phi_1^*(x,t)}{\psi_1(x,t)\phi_1^* - \phi_1(x,t)\psi_1^*} \quad (15b)$$

From the above $S_{12}$ and $S_{21}$, one should understand that the explicit forms of $\Phi_1$ which can be derived by solving equation (4) with suitable seed. With the known expression of $\Phi_1$, one can fix the matrix elements of $S_{12}$ and $S_{21}$. Finally using the $S_{12}$ and $S_{21}$ in the Darboux transformation given by (10), one can arrive the explicit soliton solution.

## IV. SOLITON SOLUTIONS

Our prior aim is to generate a more general soliton solution for the model given by equation (1). So, we omit with the vacuum seed since the bright soliton solution and its dynamics for the model have been already studied by many authors using different analytical approaches [11, 12].

To achieve more general soliton solution we start constructing the soliton using non-trivial seed. Since, constant wave background solutions are more effective one than that of the vacuum soliton solutions, because, one can able to generate the backgroundless like solutions (bright solitons) from the solitonic solutions derived using non zero seed solution. But, to generate dark solitons (background solitonic solutions) from backgroundless solitonic like (vacuum seed) solutions is impossible. Keeping all these points in mind to generate more general solution we started with constant plane wave seed. We choose $q(x, t) = Ae^{2iA^2t}, q^*(x, t) = Ae^{-2iA^2t}$ to the defocusing NLS equation ($\sigma=-1$). Inserting these seed solutions in the Lax pair equations (3) will lead to the basic solutions of the following form

$$\psi_1(x,t) = e^{iA^2t}(c1e^{\chi_{1s1}} + c2e^{-\chi_{1s1}}), \quad \ldots\ldots (16a)$$

$$\phi_1(x,t) = e^{iA^2t}\left(\frac{-c1}{A}(\lambda 1 + is_1)e^{\chi_{1s1}} + \frac{-c2}{A}(-\lambda 1 + is_1)e^{-\chi_{1s1}}\right),$$
…… (16b)

$$\psi_2(x,t) = e^{iA^2 t}(c_1^- e^{\chi_1 s_1} + c_2^- e^{-\chi_1 s_1}), \quad \ldots (16c)$$

$$\phi_2(x,t) = e^{iA^2 t}\left(\frac{c_1^-}{A}(\lambda_1 + is_1)e^{\chi_1 s_1} + \frac{-c_2^-}{A}(-\lambda_1 + is_1)e^{-\chi_1 s_1}\right) \quad \ldots \ldots (16d)$$

Where $\chi_1 = x + 2\lambda_1 t$, $\chi_2 = x + 2\lambda_2 t$, $s_1 = \sqrt{-\lambda_1^2 + A^2}$ and $s_2 = \sqrt{-\lambda_2^2 + A^2}$ Inserting these basic solutions into the Dauboux Transforms formula[10], and simplifying the resultant expression with $\tau_1 = \frac{c_1}{c_2}$ and $\tau_1 = \frac{c_1^-}{c_2^-}$ we arrive at at tedeous calculations

$$Q_1(x,t) = A e^{2iA^2 t + \theta}\left(1 + 2(\lambda_2 - \lambda_1)\frac{N_1}{D_1}\right) \quad (17a)$$

$$Q_2(x,t) = A e^{-2iA^2 t + \theta}\left(1 - \frac{2}{A^2}(\lambda_2 - \lambda_1)\frac{N_2}{D_1}\right) \quad (17b)$$

With,

$$N_1 = \tau_1 \tau_2 e^{s_1 \chi_1 + s_2 \chi_2} + \tau_2 e^{-s_1 \chi_1 + s_2 \chi_2} + \tau_1 e^{s_1 \chi_1 - s_2 \chi_2} + e^{-s_1 \chi_1 - s_2 \chi_2},$$

$$N_2 = \tau_1 \tau_2 \rho_1 \rho_2 e^{s_1 \chi_1 + s_2 \chi_2} + \tau_2 \rho_2 \rho_3 e^{-s_1 \chi_1 + s_2 \chi_2} + \tau_1 \rho_1 \rho_4 e^{s_1 \chi_1 - s_2 \chi_2} + \rho_3 \rho_4 e^{-s_1 \chi_1 - s_2 \chi_2},$$

$$D_1 = \tau_1 \tau_2 (\rho_1 - \rho_2) e^{s_1 \chi_1 + s_2 \chi_2} + \tau_3(\rho_2 - \rho_3) e^{-s_1 \chi_1 + s_2 \chi_2} + \tau_1(\rho_4 - \rho_1)e^{s_1 \chi_1 - s_2 \chi_2} - (\rho_4 - \rho_3)e^{-s_1 \chi_1 - s_2 \chi_2},$$

where $\rho_1 = \lambda_1 + is_1$, $\rho_2 = \lambda_2 + is_2$, $\rho_3 = \lambda_1\ is_1$, $\rho_4 = \lambda_2\ is_2$. This solution is more general one compared to all the solution witnessed in the literature for cubic and quintic NLS equations. While transforming this solution to model equation (1), the phase imprint parameter $\theta$ has the following form

$$\theta(x,t) = \frac{-\sqrt{\tau}(D_1 + 2N_1(\lambda_2 - \lambda_1))(a_1 a_2 D_1 + 2N_2(\lambda_1 - \lambda_2))}{D_1^2} \quad \ldots (18)$$

In the next section we have studied the dynamics of the above general solution for different parameter and displayed all types of soliton pairs using this solution.

## IV. FAMILY OF EXACT SOLUTION

The soliton solution given by Eq. (17) is the more general soliton solution for the model described by Eqs. (2). By introspection, we found all the pair of soliton solutions such as bright-bright (BB), dark-dark (DD), bright-dark (BD) and dark-bright (DB) using the above equation by finetuning the parameters suitably. The impact of external trap [13, 14], time-dependent dispersion [15] has been extensively studied from the viewpoint of Bose-Einstein condensates by one of the authors [16,17,18]. So, we demonstrated only to prove how this soliton solution is more general and how it would generate all kinds of soliton pairs and reveal under what situations it occurs with explicit parameters. We have shown all the types of soliton pairs in Fig.1 which consist of bright-dark (upper-left), dark-dark (upper-right), bright-dark (lower left) and dark-bright (lower right) for the parameters shown in the caption. By means of demonstrating all kinds of soliton pairs itself warrant its potential to act as a more general one.

FIGURE I:

Figure 1: Family of exact soliton pairs for different set of parameters as discussed in the text **(a)** upper left (bright-bright): $\lambda_1 = 0.2$, $\lambda_2 = 1$, A =3, $\tau_1 = 1 + 0.6i$, $\tau_2 = 1 − 0.6i$; **(b)** upper right (dark-dark): $\lambda_1 = −0.3$, $\lambda_2 = 0.5$, A = 3, $\tau_1 = −1 − 0.1i$, $\tau_2 = 1 + 0.1i$; **(c)** lower left (bright-dark): $\lambda_1 = 1.2$, $\lambda_2 = −1$, A = 3, $\tau_1 = 1 + 0.6i$, $\tau_2 = −0.5 + 0.3i$; **(d)** lower right (dark-bright): $\lambda_1 = −0.5$, $\lambda_2 = 1$, A = 3, $\tau_1 = −1 − 0.2i$, $\tau_2 = −0.5 − 0.3i$

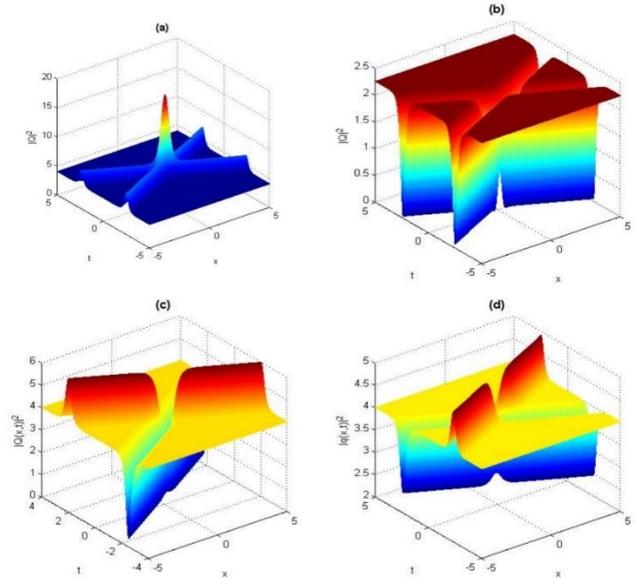

## VI. CONCLUSION

In this paper, we have investigated more general two component GP equation which represents the two component BECs at ultracold temperature. We have transformed the more general GP equation into two coupled NLS equation by using similarity transformation. We deduced more general soliton solution using Darboux and exhibited all possible soliton profiles by finetuning the parameters suitably. A refinement on the present model by incorporating PT symmetry through higher order terms is more realistic and may be tested experimentally, which we leave to future study [19].

**ACKNOWLEDGMENT**

PSV wishes to express his deepest gratitude to the Principal and Secretary of PSG College of Arts & Science for their encouragement, and he also wish to register his thanks to the Management of PSG college of Arts and Science for the SEED grant for research implementation (vide number: PSG-CAS/SGS/2019-2020/Phy/021). PSV also wishes to thank M. Venkatakrishnan and K. R. Kaviprasad for their help in editing this manuscript and fruitful discussion over the concepts involved.

**CONFLICTS OF INTEREST**

The author declares no conflicts of interest.

## REFERENCES

[1] M.H. Anderson, J.R. Ensher, M.R. Matthews, C.E. Wiemann, E.A. Cornell, "Observation of Bose-Einstein Condensation in a Dilute Atomic Vapor", *Science*, vol: 269, pp:198-201, 1995.

[2] F. Dalfovo, S. Giorgini, L.P. Pitaevskii, S. Stringari, "Theory of Bose-Einstein condensation in trapped gases", *Rev. Mod. Phys.* Vol:71, pp: 463, 1999.

[3] C. Sulem, P. L. Sulem. *The Nonlinear Schrödinger Equation*. Springer, New York, 1999.

[4] G. Theocharis, P. Schmelcher, P.G. Kevrekidis, D.J. Frantzeskakis." Matter-wave solitons of collisionally inhomogeneous condensates", *Phys. Rev. A*, vol:72, pp:033614, 2005.

[5] D.S. Wang, D.J. Zhang, J. Yang, "Integrable properties of the general coupled nonlinear Schrödinger equations", *J. Math. Phys.* Vol:51, pp:023510, 2010.

[6] D.S. Wang, X. H. Hu, J. Hu, W.M. Liu, "Quantized quasi-two-dimensional Bose-Einstein condensates with spatially modulated nonlinearity", *Phys. Rev. A*, vol:81, pp:025604, 2010.

[7] R. Radha, P. S. Vinayagam,.K. Porsezian "Soliton Dynamics of Spatially Coupled Vector BECs", *Romanian Reports in Physics,* Vol. 66, No. 2, P. 427–442, 2014.

[8] Y. H. Sabbah, Usama Al Khawaja, P. S. Vinayagam, "Lax Pair and new exact solutions of the nonlinear Dirac equation" *Com. Non. Num. Sim.* Vol: 61, pp:167-179, 2018.

[9] K. Goral, K. Rzazewski, T. Pfau, "Bose-Einstein condensation with magnetic dipole-dipole forces" *Phys Rev A*, vol:61, pp:051601, 2000.

[10] V. B. Matveev and M.A. Salle. *Darboux Transformations and Solitons*. Springer-Verlag, Berlin, 1991.

[11] R. Radha and P.S. Vinayagam. "Stabilization of matter wave solitons in weakly coupled atomic condensates", *Phys. Lett. A*, vol:376, pp:944-949, 2012.

[12] H. Chaachoua Sameut, P. Sakthivinayagam, U. Al Khawaja, Belkroukra, M. Benarous, "Peregrine Soliton Management of Breathers in Two Coupled Gross–Pitaevskii Equations with External Potential", *Phys. Wave. Phen*, vol:28, pp:305-312, 2020.

[13] R. Radha and P. S. Vinayagam, S Bhuvaneswari, R Ravisankar, P Muruganandam "Bright soliton dynamics in spin orbit-Rabi coupled Bose-Einstein condensates", *Com. Non. Num. Sim.* Vol: 50, pp:68-76, 2017.

[14] R. Radha and P.S.Vinayagam. "An Analytical Window Into The World Of Ultracold Atoms" *Romanian Reports in Physics*, Vol: 67, No. 1, pp: 89–142, 2015.

[15] R. Radha, P. S. Vinayagam, H. J. Shin and K. Porsezian, "Spatiotemporal binary interaction and designer quasi-particle condensates", *Chi. Phys. B*. vol:23(3), pp:034214, 2014.

[16] P.S Vinayagam, Amaria Javed, U Al Khawaja, "Stable discrete soliton molecules in two-dimensional waveguide arrays", *Phys. Rev. A*, vol: 98, pp:063839, 2018.

[17] R. Radha, P.S.Vinayagam, K. Porsezian, "Manipulation of light in a generalized coupled nonlinear Schrödinger equation", *Com. Non. Num. Sim.* Vol: 37, pp:354-361, 2016.

[18] U Al Khawaja, PS Vinayagam, SM Al-Marzoug. "Enhanced mobility of discrete solitons in anisotropic two-dimensional waveguide arrays with modulated separations", *Phys. Rev. A*, vol:97, pp:023820, 2018.

[19] P. S. Vinayagam, Jing Chen. "PT symmetric cubic-quintic nonlinear Schrödinger equation with dual power nonlinearities and its solitonic solutions" *Optik*. Vol:217, pp:164665, 2020.